\documentclass[aps,prb,showpacs,intlimits,amsmath,amssymb,twocolumn]{revtex4-1}
\usepackage{graphicx}
\usepackage{bm}

\begin{document}

\title{Nonequilibrium response of an electron mediated charge-density-wave-ordered material to
a large dc electric field}

\author{O.~P.~Matveev$^{1,2}$,  A.~M.~Shvaika$^2$
 T.~P.~Devereaux$^{3,4}$, and J.~K.~Freericks$^1$}
\affiliation{$^1$ Department of Physics, Georgetown University, Washington, DC
20057, USA}
\affiliation{$^2$ Institute for Condensed Matter Physics of the National Academy of Sciences of Ukraine,
Lviv, 79011 Ukraine}
\affiliation{$^3$ Geballe Laboratory for Advanced Materials, Stanford University, 
Stanford, CA 94305,USA}
\affiliation{$^4$ Stanford Institute for Materials and Energy Sciences (SIMES), 
SLAC National Accelerator Laboratory, Menlo Park, CA 94025, USA }

\begin{abstract}

Using the Kadanoff-Baym-Keldysh formalism, we 
employ nonequilibrium dynamical mean-field theory to exactly solve for the nonlinear response of an electron-mediated charge-density-wave-ordered material. We examine both the dc current and the order parameter of the conduction electrons as the ordered system is driven by the electric field. Although the formalism we develop applies to all models, for concreteness, we examine
the charge-density-wave phase of the Falicov-Kimball model, which displays a number of anomalous behaviors
including the appearance of subgap density of states as the temperature increases. These subgap states should have a significant impact on transport properties, particularly the nonlinear response of
the system to a large dc electric field.

\end{abstract}

\maketitle

\section{Introduction}
The charge-density-wave (CDW) ordered state possesses a static periodic redistribution of the electronic charge density across the 
underlying lattice. Originally, it was derived theoretically as the ground state of a one-dimensional metallic chain\cite{peierls}, where it is known as the Peierls distortion.  
Later, experiments found static CDW order occurs in materials such as the transition metal di- and trichalcogenides at nonzero temperatures.
Most of these compounds display either quasi one dimensional (NbSe$_3$) or quasi two dimensional (TaSe$_2$ or TbTe$_3$) order\cite{NbSe3,TaSe2,TbTe3}; examples of quasi three-dimensional compounds are the bismuthates BaBiO$_3$ and Ba$_{1-x}$K$_{x}$BiO$_3$~\cite{cdw_exp}.        

The CDW electronic charge redistribution is always accompanied by a lattice distortion. Even though the question for what is the fundamental principle driving the ordered phase---be it electron-driven via nesting or phonon-driven via mode softening or electron-phonon-coupling-driven---remains unsolved, recent experiments on some of these systems using time-resolved 
core-level photoemission spectroscopy\cite{hellmann} indicate an electronic nature to the ordering (in time-resolved experiments, the short-time behavior is governed by electronic relaxational processes, while the lattice responds at much longer time scales). 
This encourages us to develop the theory for the CDW phase of a strongly correlated electron material without including a direct coupling to the lattice.

Because of the extremely short relaxation times of electrons, direct experimental probes of nonequilibrium states require time resolution into the femtosecond regime or beyond.  
Recent experiments on pump-probe spectroscopy~\cite{perfetti1, perfetti2, rohwer,schmitt, petersen, hellmann} do this and display the nonequilibrium melting of the CDW state, which is 
manifested by a filling of the gap in the photoemission spectrum, while the order parameter remains nonzero.  This phenomenon has 
been theoretically examined with an exactly solvable noninteracting model~\cite{shen_fr1,shen_fr2}  starting at zero temperature. While giving insight into 
the dynamics of a driven ordered state, a full treatment of driven CDWs involving dynamic interactions is highly desired to match better with experiments.

We choose to examine the Falicov-Kimball model here.
The Falicov-Kimball model is one of the simplest models~\cite{falicov_kimball} that describes the transition to the CDW phase. It has 
an exact solution in dynamical mean-field theory\cite{brandt_mielsch1} (DMFT) in equilibrium (for a review see Ref.~\onlinecite{freericks_review}). 
For nonequilibrium, we use the Kadanoff-Baym-Keldysh formalism\cite{kadanoff,keldysh}. Since it was shown
by Langreth that many-body perturbation theory diagrams are identical for both equilibrium and nonequilibrium cases\cite{langreth},
the expansion for the local self-energy used in the equilibrium DMFT\cite{metzner} immediately extends to nonequilibrium DMFT. Hence, the lattice problem in the 
nonequilibrium case is also mapped onto an impurity problem in a time dependent field. The difference is that the dynamical mean-field now depends
on two time variables and lies on the Kadanoff-Baym-Keldysh contour. The basic structure of the iterative procedure to solve
the DMFT equations\cite{jarrell} also continues to hold. These ideas were used in recent work\cite{frtur_prb71,frturzlat_prl97,fr_prb77}
which solved the nonequilibrium DMFT for the Falicov-Kimball model in the normal phase; for a review see Ref.~\onlinecite{eckstein_rmp}. Here, we generalize this method to the case of the CDW
ordered phase.

The paper is organized as follows: in Sec.~\ref{sec:1}, we describe a formalism of the CDW phase and define a time-dependent Hamiltonian of the system.
In Sec.~\ref{sec:2}, we derive the nonequilibrium dynamical mean-field theory to calculate the lattice contour-ordered Green's function for the 
charge-density-wave system. In Sec.~\ref{sec:3}, we present our results for the nonequilibrium dc current and discussion.  We conclude in Sec.~\ref{sec:4}.

\section{CDW ordered state formalism}
\label{sec:1}

In a purely electronic theory for CDW order, the ordering arises from a nesting instability of the
Fermi surface. We examine the case of a two-sublattice CDW, which has one charge
density on the $A$ sublattice and a different charge density on the $B$ sublattice. In the ordered phase, the 
periodicity is doubled and hence the Brillouin zone (BZ) is halved (and called the reduced BZ). This scenario is described either 
by the nesting of the Brillouin zone at the modulation vector $\mathbf{Q} =(\pi, \pi, \dots)$ in reciprocal space or by introducing the two 
sublattices ``$A$'' and ``$B$'' in real space. The sublattices are defined by the modulation vector $\bf Q$ as follows
\begin{align}\label{Q}
e^{i\mathbf{Q}\cdot\mathbf{R}_i}=\begin{cases}
                                 1, & \quad \mathbf{R}_i\in A, \\
                                -1, & \quad \mathbf{R}_i\in B,
                            \end{cases} 
\end{align}
where ${\bf R}_i$ denotes the position vector for the $i$th lattice site (we work on a hypercubic lattice in the limit $d\rightarrow\infty$).
We use an additional index $\alpha=A,~B$ to denote the sublattice, so the electron annihilation
and creation operators are denoted by
\begin{equation}
 c_i^{\phantom{\dagger}}\rightarrow c_{i,\alpha}^{\phantom{\dagger}}, \quad \text{and}\quad c_i^\dagger\rightarrow c_{i,\alpha}^\dagger,\quad \alpha=A,B.
\end{equation}
To describe the ordered phase in momentum space, one has to also introduce two annihilation  and creation operators 
\begin{equation}
\tilde{c}_{1\bf k}^{\phantom{\dagger}} = c_{\bf k}^{\phantom{\dagger}}\quad \text{and} \quad \tilde{c}_{2\bf k}^{\phantom{\dagger}} = c_{\bf{k+Q}}^{\phantom{\dagger}},
\end{equation}
and 
\begin{equation}
\tilde{c}_{1\bf k}^\dagger = c_{\bf k}^\dagger\quad \text{and} \quad \tilde{c}_{2\bf k}^\dagger = c_{\bf{k+Q}}^\dagger,
\end{equation}
with the momentum $\bf{k}$ restricted to the reduced Brillouin zone (rBZ).  

In the uniform (high temperature) phase, the electron annihilation (creation) operators with momentum $\bf k$ are defined by Fourier transformation via
\begin{align}
 c_{\bf{k}}=\dfrac{1}{N}\sum_i e^{i\bf{k}\cdot\bf{R}_i}c_i,
\end{align}
with $N$ the number of lattice sites.
Similarly, exploiting the condition from Eq.~(\ref{Q}), we write down the relations between annihilation (creation) operators 
defined in the ($A,B$) sublattice basis and in the ($1,2$) rBZ basis as follows:
\begin{align}\label{relations}
\tilde{c}_{1\mathbf{k}} &=\dfrac{\sqrt{2}}{N}\sum_{i\in A}e^{i\mathbf{k}\cdot\mathbf{R}_i}c_i+\dfrac{\sqrt{2}}{N}\sum_{i\in B} e^{i\mathbf{k}\cdot\mathbf{R}_i}c_i 
= \frac{c_{\mathbf{k}A}+c_{\mathbf{k}B}}{\sqrt{2}},\\
\tilde{c}_{2\mathbf{k}} &= \dfrac{\sqrt{2}}{N}\sum_{i\in A}e^{i(\mathbf k+\mathbf Q)\cdot\mathbf{R}_i}c_i+\dfrac{\sqrt{2}}{N}\sum_{i\in B} e^{i(\mathbf k+\mathbf Q)\cdot\mathbf{R}_i}c_i \nonumber\\
&= \frac{c_{\mathbf{k}A}-c_{\mathbf{k}B}}{\sqrt{2}},
\nonumber
\end{align}
with Hermitian conjugated relations for the creation operators.
Here, the $\sqrt{2}$ factors were chosen to satisfy the standard commutation relations for the fermionic annihilation and creation operators  
\begin{equation}
[\tilde{c}_{m\mathbf k},\tilde{c}^{\dag}_{n\mathbf k'}]_{+}=\delta_{\mathbf {k,k}'}\delta_{m,n}, \qquad m,n = 1,2
\end{equation}
and
\begin{equation}
[c_{\bf k,\alpha},c^{\dag}_{\bf k',\beta}]_{+}=\delta_{\bf{k,k}'}\delta_{\alpha,\beta}, \qquad \alpha,\beta=A,B.
\end{equation}

One can rewrite the unitary transformation in Eq.~(\ref{relations}) in a matrix form as follows: 
\begin{align}
\label{transformation}
   \begin{bmatrix}
  \tilde{c}_{1\mathbf k} \\
  \tilde{c}_{2\mathbf k}
   \end{bmatrix}
=\hat{U} \begin{bmatrix}
  c_{\mathbf k A} \\
  c_{\mathbf k B}
   \end{bmatrix}, \quad\text{where}\quad
\hat{U}= \begin{Vmatrix}
  \dfrac{1}{\sqrt{2}} &  \dfrac{1}{\sqrt{2}} \\
  \dfrac{1}{\sqrt{2}} & -\dfrac{1}{\sqrt{2}}
   \end{Vmatrix}.
\end{align}
This matrix form is convenient if we need to convert from one representation to another, and we use both the  ($A,B$) sublattice
and ($1,2$) rBZ bases here. Accordingly, any two-operators-product-type quantity, {\it e.g.} the single-particle Green's function, is a $2\times2$ matrix
in the ordered state. The connection between the real space  ($A,B$) sublattice representation 
\[
\mathcal{\hat{O}}(\mathbf{k})=\|\mathcal{\hat{O}}_{\alpha,\beta}(\mathbf{k})\|,\quad \alpha,\beta=A,B,
\]
and the reciprocal space  ($1,2$) rBZ representation
\[
\mathcal{\hat{\widetilde{O}}}(\mathbf{k})=\|\mathcal{\hat{\widetilde{O}}}_{m,n}(\mathbf{k})\|, \quad m,n=1,2,
\] 
follows from the aforementioned unitary transformation via
\begin{equation}
 \mathcal{\hat{\widetilde{O}}}(\mathbf{k})=\hat{U} \mathcal{\hat{O}}(\mathbf{k}) \hat{U}^{-1}.
\end{equation}

Keeping in mind all the above descriptions of the ordered state,  we write down the time-dependent Hamiltonian of the system
\begin{equation}
\mathcal{H}(t)=\sum_{i\alpha}\mathcal{H}^\alpha_i-\sum_{ij\alpha\beta} t^{\alpha\beta}_{ij}(t)c^{\dag}_{i\alpha} c_{j\beta}^{\phantom\dagger},
\end{equation}
where the local term is equal to 
\begin{equation}
\mathcal{H}^\alpha_i=U n^\alpha_{id} n^\alpha_{if}-\mu^\alpha n^\alpha_{id},
\end{equation}
with the number operators of the itinerant and localized electrons given by 
$\hat{n}^{\alpha}_{id} = \hat{c}_{i\alpha}^\dagger \hat{c}_{i\alpha}^{\phantom\dagger}$ and $\hat{n}_{if}^{\alpha}=\hat{f}_{i\alpha}^\dagger\hat f_{i\alpha}^{\phantom\dagger}$, respectively. 
We choose different chemical potentials $\mu^A$ and $\mu^B$ for the different sublattices, which allows us to work 
with a fixed order parameter $\Delta n_f=(n_f^A-n_f^B)/2$, rather than iterating the DMFT equations to determine the order parameter (which is subject to critical 
slowing down near $T_c$\cite{fk_cdw_prb03}). The equilibrium state is achieved when the solution to the equations produces a uniform chemical potential throughout the lattice ($\mu^A=\mu^B$). We work in units 
where $\hbar=c=e=a=1$.

The system is placed into a uniform external electric field that interacts with the charged fermions. We assume that the field is spatially uniform and ignore 
all magnetic field and relativistic effects. This allows us to describe the electric field via a time-dependent vector potential in the Hamiltonian gauge:
\begin{equation}\label{field}
\mathbf{E}(t)=-\dfrac{d}{d t}\mathbf{A}(t).
\end{equation}
To describe an interaction with the external field in Eq.~(\ref{field}), we exploit a Peierls' substitution to the kinetic-energy term of the Hamiltonian. Hence, the 
Hamiltonian depends on time solely through the time-dependent vector potential as follows: 
\begin{equation}\label{hopping}
t^{\alpha\beta}_{ij}(t)=t^{\alpha\beta}_{ij}\text{exp}\biggl(-i\int\limits_{\mathbf{R}_{i,\alpha}}^{\mathbf{R}_{j,\beta}}\mathbf{A}(t)\cdot d\mathbf{r}\biggr),
\end{equation}
where $t^{\alpha\beta}_{ij}$ is the noninteracting hopping matrix.

The nonlocal kinetic-energy term in Eq.~(\ref{hopping}) of the Hamiltonian is off-diagonal in the  ($A,B$) sublattice representation since the hopping is between sites 
that belong on different sublattices for nearest-neighbor hopping. The local part of the Hamiltonian is diagonal. In the ($1,2$) rBZ representation it 
is {\it vice versa}: the nonlocal kinetic-energy part is diagonal and the local interaction $U$-term is off-diagonal.	  

In momentum space, the time-dependent kinetic-energy term has the form 
\begin{align}
\hat{\mathcal{H}}_{kin}(t) &= \sum_{k}\begin{bmatrix}
c^{\dag}_{\mathbf{k}A} & c^{\dag}_{\mathbf{k}B}
\end{bmatrix}\hat{\epsilon}(\mathbf k-\mathbf{A}(t))\begin{bmatrix}
c_{\mathbf{k}A}^{\phantom\dagger}\\ c_{\mathbf{k}B}^{\phantom\dagger}
\end{bmatrix}
\nonumber
\\
&= \sum_{k}\begin{bmatrix}
\tilde{c}^{\dag}_{1\mathbf{k}} & \tilde{c}^{\dag}_{2\mathbf{k}}
\end{bmatrix}\hat{\tilde{\epsilon}}(\mathbf k-\mathbf{A}(t))\begin{bmatrix}
\tilde{c}_{1\mathbf{k}}^{\phantom\dagger}\\ \tilde{c}_{2\mathbf{k}}^{\phantom\dagger}
\end{bmatrix},
\end{align}
where $\hat{\epsilon}(\mathbf k-\mathbf{A}(t))$ is the generalized band energy\cite{frtur_prb71} that is off-diagonal in the ($A,B$) sublattice basis
\begin{align}\label{eps}
&\hat{\epsilon}(\mathbf k-\mathbf{A}(t)) \\
&=\begin{Vmatrix}
 \scriptstyle 0 & \scriptstyle\epsilon(\mathbf k)\cos (A(t))+\bar{\epsilon}(\mathbf k)\sin (A(t)) \\
 \scriptstyle \epsilon(\mathbf k)\cos (A(t))+\bar{\epsilon}(\mathbf k)\sin (A(t)) & \scriptstyle 0
 \end{Vmatrix}
\nonumber
\end{align}
and by performing the unitary transformation in Eq.~(\ref{transformation}), we obtain an extended band energy $\hat{\tilde{\epsilon}}(\mathbf k-\mathbf{A}(t))$
in the ($1,2$) rBZ representation that is diagonal:

\begin{align}
\label{eps1}
&\hat{\tilde{\epsilon}}(\mathbf k-\mathbf{A}(t))
=\hat{U}\hat{\epsilon}(\mathbf k-\mathbf{A}(t))\hat{U}^{-1} \\
&=\begin{Vmatrix}
  \scriptstyle\epsilon(\mathbf k)\cos (A(t))+\bar{\epsilon}(\mathbf k)\sin (A(t)) & \scriptstyle 0 \\
  \scriptstyle 0 & \scriptstyle -\epsilon(\mathbf k)\cos (A(t))-\bar{\epsilon}(\mathbf k)\sin (A(t)) 
 \end{Vmatrix},
\nonumber
\end{align}
where we have chosen the electric field (and the vector potential) to lie along the diagonal (1,1,\ldots,1) direction.
Here, the generalized band energies are equal to 
\begin{align}
\epsilon(\mathbf k)&=\lim_{d\rightarrow\infty}\dfrac{-t^*}{\sqrt{d}}\sum\limits_{r=1}^d \cos k_r \quad \text{and}\nonumber \\ 
\quad \bar{\epsilon}(\mathbf k)&=\lim_{d\rightarrow\infty}\dfrac{-t^*}{\sqrt{d}}\sum\limits_{r=1}^d \sin k_r 
\end{align}
and we apply the same scaling of the hopping term as in equilibrium DMFT. 

Within DMFT, we take the limit where $d\to\infty$ and the self-energy becomes local.  It is represented by 
a diagonal matrix in the ($A,B$) sublattice representation. Alternatively, the nonequilibrium noninteracting Green's function becomes diagonal 
in the ($1,2$) rBZ basis. Below, we combine these two representations in order to obtain a self-consistent set 
of DMFT equations in nonequilibrium.

\section{Contour ordered Green's function in the CDW phase}
\label{sec:2}

In the presence of an electric field, the Hamiltonian depends explicitly on time and we must employ the Kadanoff-Baym-Keldysh formalism to describe 
the nonequilibrium behavior of the system. This approach was already used to solve for the nonlinear response of the Falicov-Kimball model\cite{fr_prb77} in the normal state. We show here how this approach generalizes to the ordered phase.

\begin{figure}
 \centerline{\includegraphics[height=0.1\textheight]{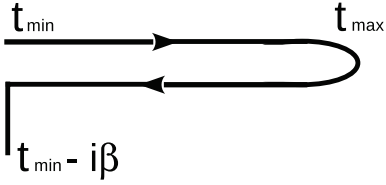}}
 \caption{Kadanoff-Baym-Keldysh time contour, which runs from a minimum time to a maximum time along the real time axis, then backwards to the minimum time, and then along the imaginary axis
for a length given by the inverse of the initial equilibrium temperature.}
 \label{contour}
\end{figure}

We begin with the contour-ordered Green's function defined on the Kadanoff-Baym-Keldysh contour in Fig.~\ref{contour}:
\begin{equation}
G^{c}_{\bf k}(t,t')=-i\langle\mathcal{T}_{c} c_{\bf k}^{\phantom\dagger}(t)c_{\bf k}^{\dag}(t')\rangle,
\end{equation}
where $\langle\mathcal{O}(t)\rangle = {\rm Tr} \exp[-\beta\mathcal{H}(t\rightarrow -\infty)]\mathcal{O}(t)/\mathcal{Z}$ and the partition function is 
$\mathcal{Z}={\rm Tr}\exp[-\beta\mathcal{H}(t\rightarrow -\infty)]$. Here, $\beta=1/T$ is the inverse of the  initial equilibrium temperature that the system started with, and we assume the Hamiltonian becomes time independent at early times. Because the momentum dependence of the Green's function depends only on the two band energies when the field is in the diagonal direction, we also use the following notation for the momentum-dependent Green's function: $G^{c}_{\bf k}(t,t')=G^c_{\epsilon,\bar{\epsilon}}(t,t')$. Note that in the above definition, the Green's function is a ``block scalar'', and it corresponds to the normal-state equation. In the ordered phase, the Green's function picks up a block $2\times 2$ matrix structure, as expected, because the momentum is ``associated'' with either the $A$ or the $B$ sublattices, and the band structure becomes an off-diagonal matrix.

Indeed, in the  ($A,B$) sublattice representation, the lattice Green's function is a $2\times2$ block matrix (denoted by the hat)
\begin{equation}
\hat{G}^c_{\epsilon,\bar{\epsilon}}(t,t')=
\begin{Vmatrix}
 G^{c,AA}_{\epsilon,\bar{\epsilon}}(t,t') & G^{c,AB}_{\epsilon,\bar{\epsilon}}(t,t') \\[1em]
 G^{c,BA}_{\epsilon,\bar{\epsilon}}(t,t') & G^{c,BB}_{\epsilon,\bar{\epsilon}}(t,t')
\end{Vmatrix},
\end{equation}
that satisfies Dyson's equation 
\begin{align}\label{deq}
&[(i{\partial_t}^c+\mu)\hat{I}-\hat{\epsilon}({\mathbf k-\mathbf{A}(t)})]\hat{G}^c_{\epsilon,\bar{\epsilon}}(t,t')\nonumber \\
&-\int_c d\bar{t}\hat{\Sigma}^c(t,\bar t)\hat{G}^c_{\epsilon,\bar{\epsilon}}(\bar{t},t')=\delta_c(t,t')\hat{I},
\end{align}
where the integral is over the contour.
The self-energy is a diagonal matrix because it is local in DMFT:
\begin{equation}
\hat{\Sigma}^c(t,\bar t)=
\begin{Vmatrix}
 \Sigma^{c,A}(t,t') & 0 \\[1em]
 0 & \Sigma^{c,B}(t,t')
\end{Vmatrix}.
\end{equation}

Here we use a short-hand notation for $\hat{\epsilon}({\mathbf k-\mathbf{A}(t)})$, which is defined in Eq.~(\ref{eps}). 
A formal solution for the lattice Green's function in Eq. (\ref{deq}) yields
\begin{equation}
\label{glat}
\hat{G}^c_{\epsilon,\bar{\epsilon}}(t,t')=\biggl{[}(\hat{G}^{c,non}_{\epsilon,\bar{\epsilon}})^{-1}-\hat{\Sigma}^c\biggr{]}^{-1}(t,t'),
\end{equation}
where the noninteracting Green's function $\hat{G}^{c,non}_{\epsilon,\bar{\epsilon}}(t,t')$ is the solution of the Dyson's
equation without interactions:
\begin{equation}\label{gnoneq}
[(i{\partial_t}^c+\mu)\hat{I}-\hat{\epsilon}({\mathbf k-e\mathbf{A}(t)})]\hat{G}^{c,non}_{\epsilon,\bar{\epsilon}}(t,t')=\delta_c(t,t')\hat{I}
\end{equation}
for the case where the field is turned on at $t=0$.

As we mentioned above, the noninteracting Green's function $\hat{G}^{c,non}_{\epsilon,\bar{\epsilon}}(t,t')$ defined by 
Eq.~(\ref{gnoneq}) is not diagonal in the ($A,B$) sublattice representation, because the generalized band energy $\hat{\epsilon}({\mathbf k-\mathbf{A}(t)})$ 
is not diagonal [see Eq.~(\ref{eps})], but the noninteracting Green's function does become block diagonal in the ($1,2$) rBZ representation:
\begin{align}
\hat{\tilde{G}}^{c,non}_{\epsilon,\bar{\epsilon}}(t,t')=
 \begin{Vmatrix}
  G^{c,non}_{\epsilon,\bar{\epsilon}}(t,t') & 0 \\[1em]
  0 & G^{c,non}_{-\epsilon,-\bar{\epsilon}}(t,t') 
 \end{Vmatrix}.
\end{align}
The analytical expression for $G^{c,non}_{\epsilon,\bar{\epsilon}}(t,t')$ is known from the uniform solution\cite{frturzlat_prl97,fr_prb77,frtur_prb71} and
is equal to:
\begin{align}
  &G^{c,non}_{\epsilon,\bar{\epsilon}}(t,t') = i[f(\epsilon-\mu)-\theta_c(t,t')]e^{i\mu(t-t')} 
 \\
 & \times e^{\left \{-i\int\limits_{t'}^{t}d\bar{t}\left ([\theta(-\bar{t})+\theta(\bar{t})\cos (A(\bar{t}))]\epsilon-\theta(\bar{t})\sin (A(\bar{t}))\bar{\epsilon}\right)\right\}}. 
 \nonumber 
\end{align}

Since the noninteracting Green's function is known, we solve for the lattice Green's function using Eq.~(\ref{glat}) in the
($1,2$) rBZ representation instead of the ($A,B$) sublattice representation. Applying the unitary transformation in Eq.~(\ref{transformation}),
we find the self-energy $\hat{\tilde{\Sigma}}^c(t,t')$ becomes: 
\begin{align}
 \hat{\tilde{\Sigma}}^c(t,\bar t)&=\hat{U}\hat{\Sigma}^c(t,\bar t)\hat{U}^{-1} 
 \\
 &=\dfrac{1}{2}
\begin{Vmatrix}
 [\Sigma^{c,A}+\Sigma^{c,B}](t,t') & [\Sigma^{c,A}-\Sigma^{c,B}](t,t') \\[1em]
 [\Sigma^{c,A}-\Sigma^{c,B}](t,t') & [\Sigma^{c,A}+\Sigma^{c,B}](t,t')
\end{Vmatrix}.\nonumber
\end{align}
Then, the solution for the lattice Green's function, in Eq.~(\ref{glat}) and in  the ($1,2$) rBZ basis, is equal to 
\begin{align}
\label{mglat}
&\hat{\tilde{G}}^c_{\epsilon,\bar{\epsilon}}(t,t')=
\begin{Vmatrix}
 \tilde{G}^{c(1,1)}_{\epsilon,\bar{\epsilon}}(t,t') & \tilde{G}^{c(1,2)}_{\epsilon,\bar{\epsilon}}(t,t') \\[1em]
 \tilde{G}^{c(2,1)}_{\epsilon,\bar{\epsilon}}(t,t') & \tilde{G}^{c(2,2)}_{\epsilon,\bar{\epsilon}}(t,t')
\end{Vmatrix} \\
&=\begin{Vmatrix}
 [\scriptstyle(G^{c,non}_{\epsilon,\bar{\epsilon}})^{-1}-\frac{\Sigma^{c,A}+\Sigma^{c,B}}{2}](t,t') & \scriptstyle-\frac{\Sigma^{c,A}-\Sigma^{c,B}}{2}(t,t') \\
 \scriptstyle-\frac{\Sigma^{c,A}-\Sigma^{c,B}}{2}(t,t') & [\scriptstyle(G^{c,non}_{-\epsilon,-\bar{\epsilon}})^{-1}-\frac{\Sigma^{c,A}+\Sigma^{c,B}}{2}](t,t')
\end{Vmatrix}^{-1}\nonumber
\end{align}

The Kadanoff-Baym-Keldysh formalism uses continuous matrix operators of two time variables defined on the contour. In computation, we need to discretize these operators into ordinary matrix operators
and then extrapolate to the continuum limit. The procedure is to solve the discretized problem with 
different time steps and then extrapolate to the continuum limit using a nonlinear extrapolation based on  Lagrange interpolation formula. Since the components in Eq.~(\ref{mglat}) are 
matrices of two time variables, we apply the block matrix pseudo-inverse formula to find the Green's function:
\begin{equation}
\label{bmi}
\begin{Vmatrix}
 A & B \\
 C & D
\end{Vmatrix}^{-1}=			
\begin{Vmatrix}
 S_D^{-1} & -A^{-1}BS_A^{-1} \\[1em]
 -D^{-1}CS_D^{-1} & S_A^{-1} 
\end{Vmatrix},
\end{equation}
where $S_A=D-CA^{-1}B$ and $S_D=A-BD^{-1}C$. 
Using this formula, we write down the components of the Green's function in the ($1,2$) rBZ basis
$\hat{\tilde{G}}^c_{\epsilon,\bar{\epsilon}}(t,t')$ as follows:
\begin{widetext}
\begin{align}
 &\tilde{G}^{c(1,1)}_{\epsilon,\bar{\epsilon}}(t,t')= 
 \biggl\{(G^{c,non}_{\epsilon,\bar{\epsilon}})^{-1}-\dfrac{\Sigma^{c,A}+\Sigma^{c,B}}{2} 
 -\dfrac{\Sigma^{c,A}-\Sigma^{c,B}}{2} \Bigl[(G^{c,non}_{-\epsilon,-\bar{\epsilon}})^{-1}-\dfrac{\Sigma^{c,A}+\Sigma^{c,B}}{2}\Bigr]^{-1}
 \dfrac{\Sigma^{c,A}-\Sigma^{c,B}}{2}\biggr\}^{-1}(t,t'), \label{eq:mom_g1} \\[1em]
 &\tilde{G}^{c(2,2)}_{\epsilon,\bar{\epsilon}}(t,t')= 
 \biggl\{(G^{c,non}_{-\epsilon,-\bar{\epsilon}})^{-1}-\dfrac{\Sigma^{c,A}+\Sigma^{c,B}}{2} 
 -\dfrac{\Sigma^{c,A}-\Sigma^{c,B}}{2}\Bigl[(G^{c,non}_{\epsilon,\bar{\epsilon}})^{-1}-\dfrac{\Sigma^{c,A}+\Sigma^{c,B}}{2}\Bigr]^{-1}
 \dfrac{\Sigma^{c,A}-\Sigma^{c,B}}{2}\biggr\}^{-1}(t,t'), %\label{eq:mom_g2}
 \end{align}
 \begin{align}\label{componentskQ}
 &\tilde{G}^{c(1,2)}_{\epsilon,\bar{\epsilon}}(t,t')=
 \biggl\{(G^{c,non}_{\epsilon,\bar{\epsilon}})^{-1}-\dfrac{\Sigma^{c,A}+\Sigma^{c,B}}{2}\biggr\}^{-1}\dfrac{\Sigma^{c,A}-\Sigma^{c,B}}{2}\tilde{G}^{c(2,2)}_{\epsilon,\bar{\epsilon}}(t,t'),
%\label{eq:mom_g3} 
\\[1em]
 &\tilde{G}^{c(2,1)}_{\epsilon,\bar{\epsilon}}(t,t')=
 \biggl\{(G^{c,non}_{-\epsilon,-\bar{\epsilon}})^{-1}-\dfrac{\Sigma^{c,A}+\Sigma^{c,B}}{2}\biggr\}^{-1}\dfrac{\Sigma^{c,A}-\Sigma^{c,B}}{2}\tilde{G}^{c(1,1)}_{\epsilon,\bar{\epsilon}}(t,t'),\label{eq:mom_g4}
\end{align}
\end{widetext}
where all inverses are with respect to the time structure of the matrices.

In the case of a CDW-ordered state,
it is more convenient to work in the  ($A,B$) sublattice basis. Hence, we apply the inverse transformation from Eq.~(\ref{transformation})
to convert from the rBZ to the sublatice representation:
\begin{equation}
\hat{G}^c_{\epsilon,\bar{\epsilon}}(t,t')=\hat{U}^{-1}\hat{\tilde{G}}^c_{\epsilon,\bar{\epsilon}}(t,t')\hat{U}.
\end{equation}
Then, we obtain the new Green's function  as follows: 
\begin{align}\label{componentsAB1}
G^{c(A,A)}_{\epsilon,\bar{\epsilon}}(t,t')&=\dfrac{1}{2}\Bigl(\tilde{G}^{c(1,1)}_{\epsilon,\bar{\epsilon}}(t,t')+\tilde{G}^{c(2,2)}_{\epsilon,\bar{\epsilon}}(t,t')\nonumber \\
                                                          &+\tilde{G}^{c(1,2)}_{\epsilon,\bar{\epsilon}}(t,t')+\tilde{G}^{c(2,1)}_{\epsilon,\bar{\epsilon}}(t,t') \Bigr), \\
\label{componentsAB2}
G^{c(B,B)}_{\epsilon,\bar{\epsilon}}(t,t')&=\dfrac{1}{2}\Bigl(\tilde{G}^{c(1,1)}_{\epsilon,\bar{\epsilon}}(t,t')+\tilde{G}^{c(2,2)}_{\epsilon,\bar{\epsilon}}(t,t')\nonumber \\
                                                          &-\tilde{G}^{c(1,2)}_{\epsilon,\bar{\epsilon}}(t,t')-\tilde{G}^{c(2,1)}_{\epsilon,\bar{\epsilon}}(t,t') \Bigr), \\
\label{componentsAB3}
G^{c(A,B)}_{\epsilon,\bar{\epsilon}}(t,t')&=\dfrac{1}{2}\Bigl(\tilde{G}^{c(1,1)}_{\epsilon,\bar{\epsilon}}(t,t')-\tilde{G}^{c(2,2)}_{\epsilon,\bar{\epsilon}}(t,t')\nonumber \\
                                                          &+\tilde{G}^{c(2,1)}_{\epsilon,\bar{\epsilon}}(t,t')-\tilde{G}^{c(1,2)}_{\epsilon,\bar{\epsilon}}(t,t') \Bigr),\\
\label{componentsAB4}
G^{c(B,A)}_{\epsilon,\bar{\epsilon}}(t,t')&=\dfrac{1}{2}\Bigl(\tilde{G}^{c(1,1)}_{\epsilon,\bar{\epsilon}}(t,t')-\tilde{G}^{c(2,2)}_{\epsilon,\bar{\epsilon}}(t,t')\nonumber \\
                                                          &+\tilde{G}^{c(1,2)}_{\epsilon,\bar{\epsilon}}(t,t')-\tilde{G}^{c(2,1)}_{\epsilon,\bar{\epsilon}}(t,t') \Bigr),
\end{align}
using the results from Eqs.~(\ref{eq:mom_g1}--\ref{eq:mom_g4}).

In DMFT, we need to solve the many-body problem on an impurity which mimics the behavior of the lattice (and yields the same local self-energy). This requires us to map the lattice problem onto an impurity problem. To do so requires us to first determine the local Green's function
by summing the $\epsilon,\bar\epsilon$-dependent functions over the generalized band energies weighted by the appropriate joint density of states. This must be done for the two local Green's functions (one for each of the $A$ and $B$ sublattices) and becomes a double integral given by
\begin{align}\label{glocsi}
 \hat{G}^c_{loc}(t,t')&=
  \int d\epsilon\int d\bar{\epsilon} \rho(\epsilon,\bar{\epsilon})\hat{G}^c_{\epsilon,\bar{\epsilon}}(t,t')\nonumber \\[1em]
 &=\begin{Vmatrix}
  G^{c,A}_{loc}(t,t') & 0 \\[1em]
  0 & G^{c,B}_{loc}(t,t')
 \end{Vmatrix},
\end{align}
where the joint density of states is a Gaussian in each variable for the hypercubic lattice \cite{frtur_prb71} given by
\[
\rho(\epsilon,\bar{\epsilon})=\dfrac{1}{\pi}e^{-\epsilon^{2}-\bar{\epsilon}^{2}}.
\]

Substituting the components from Eqs.~(\ref{componentsAB1}--\ref{componentsAB4}) into Eq.~(\ref{glocsi}), we obtain an expression for the
components of the local Green's function for each sublattice as follows:
\begin{align}\label{gloc}
&G^{c,A}_{loc}(t,t')
  =\int \int d\epsilon d\bar{\epsilon} \rho(\epsilon,\bar{\epsilon})[I+\Lambda\Sigma][I-K\Sigma\Lambda\Sigma]^{-1}K,\nonumber \\
&G^{c,B}_{loc}(t,t')
 =\int \int d\epsilon d\bar{\epsilon} \rho(\epsilon,\bar{\epsilon})[I-\Lambda\Sigma][I-K\Sigma\Lambda\Sigma]^{-1}K, 
\end{align}
where we introduced short-hand notations for $K$, $\Lambda$, and $\Sigma$, which satisfy
\begin{align}\label{kls}
K&= 
 \biggl\{(G^{c,non}_{\epsilon,\bar{\epsilon}})^{-1}-\dfrac{\Sigma^{c,A}+\Sigma^{c,B}}{2}\biggr\}^{-1}(t,t'), \nonumber \\
\Lambda&= 
 \biggl\{(G^{c,non}_{-\epsilon,-\bar{\epsilon}})^{-1}-\dfrac{\Sigma^{c,A}+\Sigma^{c,B}}{2}\biggr\}^{-1}(t,t'), \nonumber \\
\Sigma&=
 \dfrac{\Sigma^{c,A}-\Sigma^{c,B}}{2}(t,t'),
\end{align}
When there is no CDW order, we have $\Sigma^{c,A}(t,t')=\Sigma^{c,B}(t,t')=\Sigma^{c}(t,t')$, and
the above formulas  reduce to the known result for the uniform phase $G^{c,A}_{loc}(t,t')=G^{c,B}_{loc}(t,t')=G^{c}_{loc}(t,t')
=\int \int d\epsilon d\bar{\epsilon} \rho(\epsilon,\bar{\epsilon})
\bigl\{(G^{c,non}_{\epsilon,\bar{\epsilon}})^{-1}(t,t')-\Sigma^c(t,t')\bigr\}^{-1}$.

In order to determine and then solve the associated impurity problem, we first exploit Dyson's equation to extract an effective medium
$\hat{G}^{c}_0(t,t')$ (all terms are block $2\times 2$ matrices)
\begin{equation}\label{g0}
 \hat{G}^{c}_{loc}(t,t')=[(\hat{G}^{c}_0(t,t'))^{-1}(t,t')-\hat{\Sigma}^{c}(t,t')]^{-1}=\hat{G}^{c}_{imp}(t,t'). 
\end{equation}
In the  ($A,B$) sublattice representation, the effective medium $\hat{G}^{c}_0(t,t')$ is block diagonal and its components are equal to 
\begin{align}
 &G^{c,A}_0(t,t')=[(G^{c,A}_{loc})^{-1}+\Sigma^{c,A}]^{-1}(t,t'), \nonumber \\
 &G^{c,B}_0(t,t')=[(G^{c,B}_{loc})^{-1}+\Sigma^{c,B}]^{-1}(t,t').
\end{align}
The effective medium  can be also found from Dyson's equation, which 
defines an effective dynamical mean field $\hat{\lambda}^{c}(t,t')$ via
\begin{equation}
\label{lambdag0}
(i{\partial_t}^c+\mu)\hat{G}^c_{0}(t,t')
-\int_c d\bar{t}\hat{\lambda}^c(t,\bar t)\hat{G}^c_{0}(\bar{t},t')=\delta_c(t,t')\hat{I}.
\end{equation}
The effective medium is then expressed as follows
\begin{align}
&G^{c,A}_{0}(t,t')=[(i{\partial_t}^c+\mu)\delta_c(t,t')-\lambda^{c,A}(t,t')]^{-1}, \nonumber \\
&G^{c,B}_{0}(t,t')=[(i{\partial_t}^c+\mu)\delta_c(t,t')-\lambda^{c,B}(t,t')]^{-1},
\end{align}
in terms of the dynamical mean fields.
Inverting these relations allows us to determine the dynamical mean fields directly
\begin{align}\label{lambda}
\lambda^{c,A}(t,t')&=(i{\partial_t}^c+\mu)\delta_c(t,t')-(G^{c,A}_{0})^{-1}(t,t') \nonumber \\
&=(i{\partial_t}^c+\mu)\delta_c(t,t')-(G^{c,A}_{loc})^{-1}(t,t')-\Sigma^{c,A}(t,t'), \nonumber \\
\lambda^{c,B}(t,t')&=(i{\partial_t}^c+\mu)\delta_c(t,t')-(G^{c,B}_{0})^{-1}(t,t') \nonumber \\
&=(i{\partial_t}^c+\mu)\delta_c(t,t')-(G^{c,B}_{loc})^{-1}(t,t')-\Sigma^{c,B}(t,t'),
\end{align}
which are the effective time-dependent fields for the associated nonequilibrium single-impurity problems.

The final step is to solve each impurity problem for the impurity Green's function in terms of the
dynamical mean field.  The result, for the Falicov-Kimball model is
\begin{align}\label{gimp}
 &G^{c,A}_{imp}(t,t')=(1-n_f^A)G^{c,A}_0(t,t')+n_f^AG^{c,A}_1(t,t'), \nonumber \\
 &G^{c,B}_{imp}(t,t')=(1-n_f^B)G^{c,B}_0(t,t')+n_f^BG^{c,B}_1(t,t'),
\end{align}
where
\begin{align}\label{g1}
 G^{c,\alpha}_1(t,t')=[1-G^{c,\alpha}_0(t,t')U]^{-1}G^{c,\alpha}_0(t,t'),\quad \alpha=A,B,
\end{align}
and $n_f^\alpha$ is the density of the heavy electrons on each sublattice (which is set by the equilibrium value and does not change when the field is applied because it does not directly
couple to the field).

In summary, nonequilibrium DMFT algorithm is as follows: First, we solve the equilibrium problem to determine $n_f^\alpha$ and $\mu$ for the given values of the 
interaction parameters and the temperature. This then determines the order parameter value $\Delta n_f=(n_f^A-n_f^B)/2$ (which runs from 0 at $T_c$ to 0.5 at $T=0$). Next, we use the equilibrium 
results for $\Sigma_{eq}^A$ and $\Sigma_{eq}^B$ as the initial guess for the nonequilibrium self-energy, we determine
$G^{c,non}_{\epsilon,\bar{\epsilon}}(t,t')$ and $G^{c,non}_{-\epsilon,-\bar{\epsilon}}(t,t')$, and then calculate  the local Green's functions 
$G^{c,A}_{loc}(t,t')$ and $G^{c,B}_{loc}(t,t')$ as in Eq.~(\ref{gloc}). Third, we extract the dynamical mean fields 
$\lambda^{c,A}(t,t')$ and $\lambda^{c,B}(t,t')$ by employing the Dyson equation as in Eq.~(\ref{lambda}). Then we calculate the effective mediums $G^{c,A}_{0}(t,t')$, $G^{c,B}_{0}(t,t')$ 
from Eq.~(\ref{lambdag0}) and $G^{c,A}_{1}(t,t')$, $G^{c,B}_{1}(t,t')$ from Eq.~(\ref{g1}). Fourth, we solve for the impurity Green's functions 
$G^{c,A}_{imp}(t,t')$ and $G^{c,B}_{imp}(t,t')$ from Eq.~(\ref{gimp}). Finally, we set $\hat{G}^{c}_{loc}(t,t')=\hat{G}^{c,A}_{imp}(t,t')$ 
and extract a new self-energy $\hat{\Sigma}^{c}(t,t')$ from Eq.~(\ref{lambda}). We repeat these steps iteratively until the solutions have converged to the desired accuracy. Note that this procedure describes how the problem is solved for a particular discretization of the Kadanaoff-Baym-Keldysh contour. 
Using a range of different discretizations, we then extrapolate the results of at least three different discretizations to determine the continuum 
limit (we demonstrate this in Ref.~\onlinecite{noneq_cdw_arxiv15}).

The formalism developed here is a fully general formalism that can be applied to other models that also display CDW order, like the attractive Hubbard model or the Holstein model. The only difference is how the impurity Green's function is determined from the effective medium, which is given here explicitly for the Falicov-Kimball model, but needs to be solved approximately for other models (or with more sophisticated numerical techniques).

As a check of our numerical results, we have verified the spectral moment sum rules which continue to hold in ordered phase and in nonequilibrium. The moments are defined to be
\begin{eqnarray}
&~& \mu_{n}^{R,a}(t_{ave})=\\
&~&-\dfrac{1}{\pi}\int\limits_{-\infty}^{\infty}d\omega\text{Im}\int\limits_{-\infty}^{\infty}dt_{rel} e^{i\omega t_{rel}}i^{n}\dfrac{\partial^n}{\partial t^n}G^{R,\alpha}(t_{ave},t_{rel}),\nonumber
\end{eqnarray}
with $\alpha=(A,B)$ and the retarded Green's function is expressed in terms of the Wigner coordinates $t_{ave}=(t+t')/2$ and $t_{rel}=t-t'$. We have calculated the zeroth, first and second moments 
which satisfy\cite{frtur_sumrules} 
\begin{equation}
 \mu_{0}^{R,\alpha}(T)=1, 
\end{equation}
\begin{equation}
 \mu_{1}^{R,\alpha}(T)=-\mu+Un_{f}^{\alpha}, 
\end{equation}
\begin{equation}
 \mu_{2}^{R,\alpha}(T)=\dfrac{1}{2}+\mu^{2}-2U\mu n_{f}^{\alpha}+U^{2}n_{f}^{\alpha},
\end{equation}
where $n_{f}^{A}=1/2+\Delta n_{f}$ and $n_{f}^{B}=1/2-\Delta n_{f}$. For different time discretizations, higher accuracy occurs for larger average time and smaller
interaction $U$. We find the results for the extrapolated moments agree with the exact results to high accuracy for all parameters presented in this work. 

We end with a discussion of the order parameters. There are two different ones that can be chosen. One is the order parameter of the heavy electrons, 
which we introduced above and it is the difference of the heavy electron filling on the $A$ and $B$ sublattices. The order parameter starts at 0 at $T_c$ and increases 
all the way to 0.5 at $T=0$. This order parameter remains unchanged as the field is applied, because the heavy electrons do not couple to the external electric field since they are localized. The other order parameter is the difference in the conduction electron filling on the two sublattices. This order parameter can change as the field is applied, because the field does cause motion of the conduction electrons. Even in equilibrium, this order parameter is distinct from the heavy electron order parameter; it starts at 0 at $T_c$ but it generically does not go to 1 as $T\rightarrow 0$. Indeed, it can even change sign when the system is pumped by an external electric field.

\section{Results}
\label{sec:3}

We show our results for the concentration of the conduction electrons and for the current in nonequilibrium when a uniform $dc$ electric field is suddenly switched on at time $t=0$.
The conduction electron filling is determined by the imaginary part of the lesser Green's function $G_{\mathbf{k}}^<(t,t)$ at equal times, which is extracted from the contour-ordered Green's function. The current is calculated in the Hamiltonian gauge by evaluating the operator average
\begin{align}
 \langle\mathbf{j}(t)\rangle=-i\sum\limits_{\mathbf{k}}\mathbf{v}(\mathbf{k}+\theta(t)\mathbf{E}t)G_{\mathbf{k}}^<(t,t), 
\end{align}
where the velocity component is $v_i(\mathbf{k})=\lim\limits_{d\rightarrow\infty} t^{*}\sin(\mathbf{k}_i)/\sqrt{d}$ (note that the final result for the current, which is an observable, is gauge invariant).  
The field magnitude is set to $E=1$.

\begin{figure}[htb]
\centerline{ \includegraphics[width=0.55\textwidth]{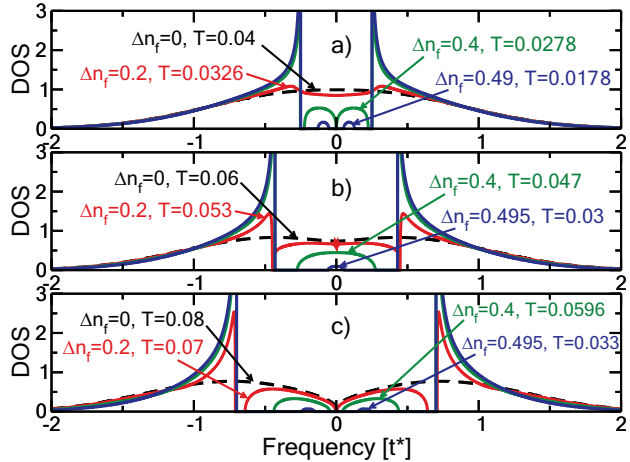}}
 \caption{(Color online.) Equilibrium density of states for different values of interaction $U$: a) $U=0.5$ corresponds to a metal in the normal state and 
 an insulator in the ordered phase ($T_c=0.0336$); b) $U=0.86$ corresponds to a metal in the normal state and a quantum-critical case in the ordered phase ($T_c=0.055$); 
  c) $U=1.4$ corresponds to a semiconductor in normal state and a strongly correlated insulator in the ordered phase ($T_c=0.0727$). Different curves correspond to different temperatures as marked by the arrows.}
 \label{eqdos}
\end{figure}
In Fig.~\ref{eqdos}, we plot the equilibrium density of states (DOS) for different values of the interaction  $U$ and for different temperatures. By decreasing the temperature,
we push system into the CDW ordered phase with a gap in the DOS, which equals  $U$ at zero temperature. For nonzero temperatures, the gap is partially filled with subgap states
which can affect transport properties of the system in equilibrium~\cite{msf_opt}. Here we are interested in how these peculiarities of the equilibrium DOS are modified in the
nonequilibrium case. 

\begin{figure}[htb]
 \centerline{\includegraphics[width=0.49\textwidth]{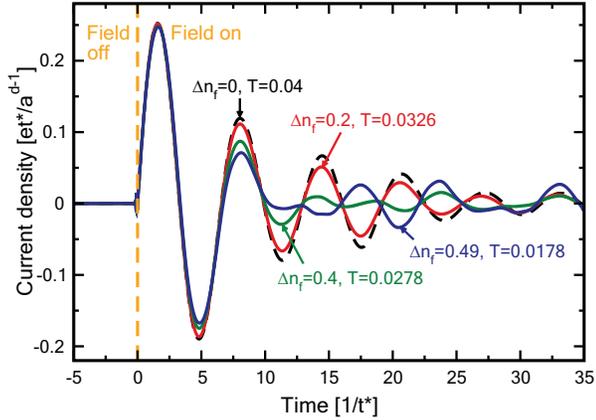}}
 \caption{(Color online.) Current for $U=0.5$ ($T_c=0.0336$) with an electric field $E=1$. Different curves correspond to different initial temperatures and correspondingly different initial order parameters.}
 \label{curru05}
\end{figure}
\begin{figure}[htb]
\centerline{ \includegraphics[width=0.49\textwidth]{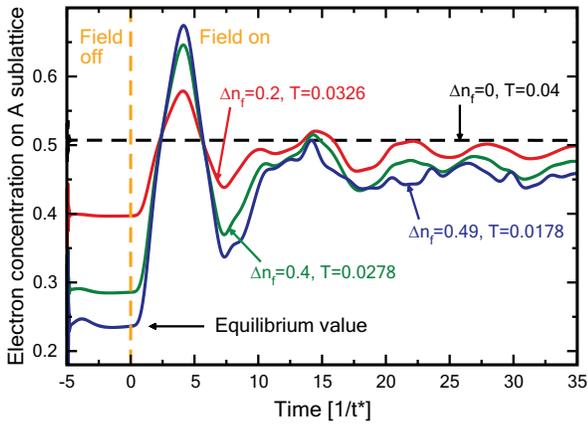}}
 \caption{(Color online.) Concentration of conduction electrons on the $A$ sublattice for $U=0.5$ ($T_c=0.0336$) with an electric field $E=1$. 
 Different curves correspond to different initial temperatures.}
 \label{concu05}
\end{figure}

In Fig.~\ref{curru05}, we plot the current at $U=0.5$, which corresponds to a metal in the normal state, and a weakly correlated CDW insulator in the ordered phase. Different curves
correspond to different temperatures starting from above the critical temperature ($T_c=0.0336$) and running down to $T=0.0178$, where the order parameter is nearly maximal ($\Delta n_f=0.49$). If there is no interaction and the external electric field is constant, we expect a
permanently oscillating current called a Bloch oscillation. But in the case of an interacting system, the oscillations are damped at long times.
In the absence of order, the current amplitude decays relatively slowly but as the order parameter increases ($\Delta n_f=0.2$), the current amplitude decays faster because of the presence of a gap in the equilibrium
single-particle density of states. But as the temperature is lowered further, and one of the sublattices becomes almost fully occupied by the $f$-electrons (and other becomes nearly empty),
the oscillations in the current become long-lived since the scattering is sharply reduced (and ultimately vanishes at $T=0$). This low-temperature behavior agrees well with the zero temperature results solved with a completely different method\cite{shen_fr2}. 

Fig.~\ref{concu05} plots the conduction-electron filling for the same parameters.
When the field is initially turned on, the value of the electron filling changes dramatically (even to the point of inverting the conduction electron CDW order parameter) but for longer times it settles into a reduced order parameter,
which has complex oscillations. The conduction electron order parameter can be read off of this plot
because $\rho_A+\rho_B=1$ and the order parameter is $\rho_A-\rho_B=2\rho_A-1$. Hence the order parameter changes sign when the conduction electron density on the $A$ subattice crosses 0.5.

\begin{figure}[htb]
\centerline{ \includegraphics[width=0.49\textwidth]{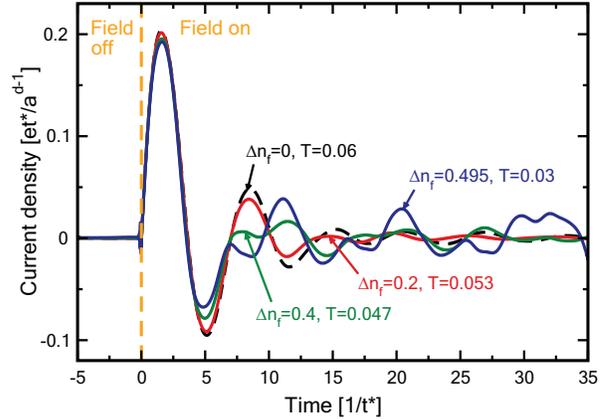}}
 \caption{(Color online.) Current for $U=0.86$ ($T_c=0.055$) with electric field $E=1$. Different curves correspond to different initial temperatures.}
 \label{curru086}
\end{figure}

\begin{figure}[htb]
\centerline{ \includegraphics[width=0.49\textwidth]{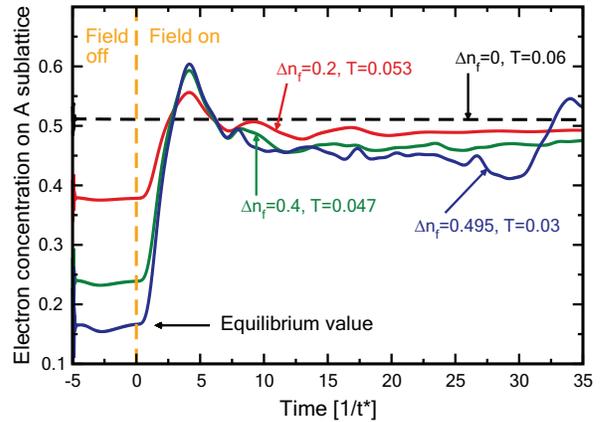}}
 \caption{(Color online.) Conduction electron filling on the $A$ sublattice for $U=0.86$ ($T_c=0.055$) with electric field $E=1$. 
 Different curves correspond to different initial temperatures.}
 \label{concu086}
\end{figure}

Next, we examine the quantum-critical case with interaction $U=0.86$ in Figs.~\ref{curru086} and \ref{concu086}. The CDW for this $U$ value is in the quantum critical metallic CDW phase,
where the density of states fills in at $\omega=0$ for all finite $T$ and hence will have metallic conduction for all finite $T$. The current in Fig.~\ref{curru086} behaves similar 
to the previous case of $U=0.5$, since the system is also a metal in the uniform phase. Two differences appear in the figures:  the current has a lower magnitude and approaches the steady-state faster. For the filling in Fig.~\ref{concu086}, we see a similar scenario, 
namely the concentration of conduction electrons reaches the steady state more rapidly. 
The lowest temperature case shows an odd reversal of the order parameter at long times as well.

\begin{figure}[htb]
\centerline{ \includegraphics[width=0.49\textwidth]{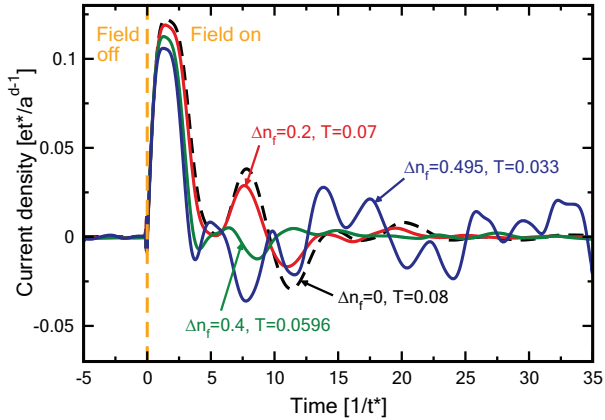}}
 \caption{(Color online.) Current for $U=1.4$ ($T_c=0.0727$) with electric field $E=1$. Different curves correspond to different temperatures.}
 \label{curru14}
\end{figure}

\begin{figure}[htb]
\centerline{ \includegraphics[width=0.49\textwidth]{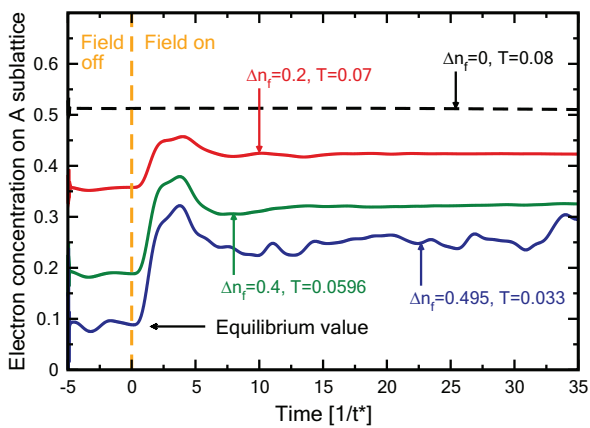}}
 \caption{(Color online.) Filling of conduction electrons on the $A$ sublattice for $U=1.4$ ($T_c=0.0727$) with electric field $E=1$. 
 Different curves correspond to different initial temperatures.}
 \label{concu14}
\end{figure}

In Fig.\ref{curru14}, we show the results for the current for $U=1.4(\approx\sqrt{2})$ which is at the metal-insulator transition and corresponds to the strongly correlated CDW. Because of the fact that the system is at the metal-insulator transition, the amplitude of the current, even above the critical temperature, is half the size of the metallic case with $U=0.5$. This is due to the fact that in the uniform phase, the Mott gap starts to develop at 
$U=\sqrt{2}$, so the normal-state density of states already shows a pseudogap (see Fig.~\ref{eqdos}). Additionally, the 
oscillations damp faster. As the temperature dips below the critical temperature, these effects are initially enhanced due to the CDW gap formation. Lowering the
temperature further, starts to see the damping disappear as the ordered phase has less and less scattering as $T\rightarrow 0$. 

In Fig.~\ref{concu14}, we show results for the  filling of the conduction electrons on the $A$ sublattice.
Once the field is turned on, the conduction electron filling is sharply changed, but not enough to reverse the order as it did previously.  The damping of the oscillations is very rapid when the order parameter is small and there is significant scattering. When the order parameter gets large, the oscillations survive for a long time and are irregular. At the longest times, the system approaches a steady state, similar to the one seen
at zero temperature\cite{shen_fr2}.

One of the interesting observations that can be seen in this data is that the weakly correlated CDW states seem to nearly lose the CDW order parameter for the 
conduction electrons very easily, while in the insulating phase, the order remains more robust, and does not get reduced as rapidly or as much (compare 
Fig.~\ref{concu05} to Fig.~\ref{concu14}). Another is that the oscillatory behavior in the current often becomes out of phase as the order parameter increases 
at lower teperatures. In this model, the system becomes noninteracting at zero temperature, because all scattering is suppressed. Nevertheless, complex time 
traces develop when one is nearly fully ordered, which are indicative of the behavior seen in Mott insulators in the normal phase (such as the irregular 
oscillations in the current or in the order parameter). It is curious that such behavior survives, even as the system has less and less scattering due to 
the increased order.

\section{Conclusions}
\label{sec:4}

In this work, we described the general formalism for how to solve nonequilibrium DMFT in an ordered CDW phase. For concreteness we chose to examine the Falicov-Kimball model. We derived analytical expressions for the time-dependent lattice Green's functions
defined on the contour by generalizing the results for the paramagnetic phase\cite{fr_prb77,frtur_prb71}   to
the  two-sublattice case. We studied the simplest time dependence of the external field with a constant field $E=1$ turned on at $t=0$. Of course, other temporal field profiles can also be used within this 
formalism.

We showed how the current and the filling of the conduction electrons behave in the CDW phase when the system is driven into nonequilibrium by a large $dc$ electric field. We examined cases with different interactions
$U$ corresponding to a metallic phase ($U=0.5$), to a quantum critical point ($U=0.86$) and to a critical Mott insulator ($U=1.4$). The Bloch oscillations of the 
current, which decreased smoothly in the normal state, initially vanish more rapidly in the  CDW phase. As $T\rightarrow 0$, the current demonstrates complex 
oscillations  and indicates the possible formation of a steady state.

\begin{acknowledgments}
This work was supported by the Department of Energy, Office of Basic Energy Sciences, Division of Materials Sciences and Engineering under Contract Nos. DE-AC02-76SF00515 
(Stanford/SIMES), DE-FG02-08ER46542 (Georgetown) and DE-SC0007091 (for the collaboration). Computational resources were provided by the National Energy Research Scientific 
Computing Center supported by the Department of Energy, Office of Science, under Contract No. DE- AC02-05CH11231. J.K.F. was also supported by the McDevitt Bequest at Georgetown. 
\end{acknowledgments}

\end{document}